\documentclass[]{elsart5p}
\usepackage[normalem]{ulem}
\usepackage{hyperref}
\usepackage[margin=10pt, font=small, labelfont=bf]{caption}   
\usepackage{graphicx}        
\usepackage{float}
\usepackage{booktabs}               
\usepackage{natbib}                  
\usepackage[page]{appendix}

\begin{document}

\begin{frontmatter}
\title{\small This paper has been published in Biological Conservation (2008) 141, 2625-2631, doi:10.1016/j.biocon.2008.07.028\\[0.4cm]
\LARGE The time horizon and its role in multiple species conservation planning}
\journal{Biological Conservation}
\author[UFZ]{Florian HARTIG\corauthref{cor}}
\corauth[cor]{Corresponding author, Tel: +49-341-235-1716, Fax: +49-341-235-1473}
\ead{florian.hartig@ufz.de}
\author[UFZ]{Martin DRECHSLER}
\ead{martin.drechsler@ufz.de}

\address[UFZ]{UFZ - Helmholtz Centre for Environmental Research, Department of Ecological Modelling, Permoserstr. 15, 04318 Leipzig, Germany}
\begin{abstract}
Survival probability within a certain time horizon $T$ is a common measure of population viability. The choice of $T$ implicitly involves a time preference, similar to economic discounting: Conservation success is evaluated at the time horizon $T$, while all effects that occur later than $T$ are not considered. Despite the obvious relevance of the time horizon, ecological studies seldom analyze its impact on the evaluation of conservation options. In this paper, we show that, while the choice of $T$ does not change the ranking of conservation options for single species under stationary conditions, it may substantially change conservation decisions for multiple species. We conclude that it is of crucial importance to investigate the sensitivity of model results to the choice of the time horizon or other measures of time preference when prioritizing biodiversity conservation efforts.
\end{abstract}

\end{frontmatter}

\section{Introduction}
A central problem for conservation planning is the decision on conservation goals \citep{Margules-Systematicconservationplanning-2000}. These goals are used to define quantitative objective functions, which are needed for a systematic comparison of conservation options \citep[see e.g.][]{Wilson-Prioritizingglobalconservation-2006}. Depending on societal preferences, a number of ecosystem properties and services may be valued, and objective functions may vary accordingly \citep{Balvanera-ConservingBiodiversityand-2001, Williams-Applesorangesand-2002, Roberts-Applicationofecological-2003}. 
For conservation planning, objectives built on measures such as percentage of preserved area or expected coverage are traditionally used because they are relatively easy to apply; however, it has been repeatedly shown that these measures may fail to act as a reliable surrogate for the persistence of species \citep{Cabeza-Site-selectionalgorithmsand-2003, Svancara-Policy-drivenversusevidence-based-2005, Wiersma-Conservationtargetsviable-2006}. Species survival probabilities, in contrast, provide a measure which relates directly to the actual goal of persistence and thus acts as a better predictor for conservation success \citep{Williams-Usingprobabilityof-2000, Guisan-Predictingspeciesdistribution-2005}.\\\\
Translating the goal of persistence into a quantitative objective based on survival probabilities needs some further clarification when dealing with multiple species. A number of different objective functions are used in the literature. Some maximize the expected number of surviving species, others use the probability of all species surviving, or the probability of the most threatened species surviving \citep[see e.g.][]{Bevers-Sustainableforestmanagement-1995, Nicholson-Objectivesmultiple-speciesconservation-2006}. Although all aiming at improving species survival, these objectives may vary substantially in their rating of conservation options and subsequently in their choice of conservation priorities \citep{Nicholson-Objectivesmultiple-speciesconservation-2006}.\\\\
Despite their differences, all these functions express survival in terms of the probability of surviving until some time $T$. $T$, frequently called the time horizon or the time frame, is a preferred time at which conservation success is evaluated. A short time horizon acts similarly to a large discounting factor in economics and vice versa. While the choice of such time preferences is subject to serious debate in the field of environmental economics \citep{Rabl-Discountingoflong-term-1996, Weitzman-WhyFar-DistantFuture-1998, Heal-Discountingreviewof-2007}, it seems that conservation planning has widely neglected this topic so far. Some of the rare exceptions include \citet{Eiswerth-Maximizingconservedbiodiversity-2001} and \citet{Cabeza-Site-selectionalgorithmsand-2003}. One explanation may be that the time horizon usually has no impact on static single-species conservation, and it is believed that the same holds true for multi-species conservation. Another reason may be that the controversy about discounting ecological values has been considered a social science issue much more than an ecological question. Nevertheless, our results show that excluding this discussion from the scope of conservation planning may result in misleading and possibly unintended conservation recommendations. \\\\
In this paper, we analyze three typical objective functions which are used in the literature with respect to their sensitivity to the choice of the time horizon. We find that, for additive functions, this choice may have a crucial impact on the resulting conservation decisions. We conclude that the choice of a time horizon is an inevitable part of decision making. Its influence must be borne in mind and should be explicitly communicated when determining conservation targets.   
\section{Methods and assumptions}
\subsection{The time horizon and annual survival}
Under stationary environmental conditions (no trends in population parameters such as carrying capacity, so that the population is in a quasi-stationary state), the probability of surviving until time $T$ is given by
\begin{equation} \label{eq: meantimetoext}
    p(T) = e^{-\frac{T}{T_{m} }}
\end{equation} 
where $T_m$ is the mean time to extinction \citep{Grimm-intrinsicmeantime-2004}, measured in years. The annual survival probability is $x = exp(-1/T_m)$. With eq. \ref{eq: meantimetoext}, we can then express the survival of a species until time $T$ by    
\begin{equation} \label{eq: expdecay}
     p(T)  =  x^T
\end{equation}
where $x$ denotes the annual survival probability as given before. Using this as the basis of our evaluation, we should first note a trivial, but crucial fact: The survival probability $p$ decreases nonlinearly (exponentially) with the time horizon $T$. For a stationary single species case under stationary external conditions, however, this nonlinearity does not change ratings based on the survival probability $p$; given that a conservation option has a higher $p(T_0)$ than another option for a time horizon $T_0$, it will also have a higher $p(T)$ for any other time horizon $T$. \\\\

\subsection{Multi-species objective functions}
For the case of multiple species, knowledge of single species survival probabilities is not enough to compare conservation options. As an example, imagine the case of two species, and two conservation alternatives, one which yields survival probabilities of $p_1=70\%$ and $p_2 = 90\%$, and another which results in $p_1=80\%$ and  $p_2 = 80\%$. Which option is to be preferred? The expectation value of the number of species surviving, $p_1 + p_2$, is the same for both cases. Yet, the second conservation alternative shows a more even distribution of survival probabilities between species. \\\\ 
The literature has approached the problem of multi-species survival mainly with two classes of objective functions: additive and multiplicative ones \citep[see][]{Nicholson-Objectivesmultiple-speciesconservation-2006}. In its most simple form, an additive objective function for $n$ species is given by the sum of the single species survival probabilities $p_i$:
\begin{equation}\label{eq: additive}
    \sum_{i=1}^n p_i\; 
\end{equation}
Mathematically, the sum represents the expected value of the number of species surviving. Examples of studies using additive functions are \citet{Faith-Integratingconservationand-1996, Polasky-Selectingbiologicalreserves-2001, Nicholson-newmethodconservation-2006}. A simple multiplicative function is given by the product of all survival probabilities:
\begin{equation}\label{eq: multiplic}
    \prod_{i=1}^n p_i 
\end{equation}
This product represents the probability that all species survive \citep[see e.g.][]{Bevers-Sustainableforestmanagement-1995}. Multiplicative objective functions tend to favor an even distribution of survival probabilities, whereas additive objectives generally do not \citep[][]{Nicholson-Objectivesmultiple-speciesconservation-2006}. In the context of biodiversity, such an evenness objective is often considered advantageous. However, it is also possible to include evenness objectives in additive objective functions \citep[see e.g.][]{Arponen-valueofbiodiversity-2005, Moilanen-LandscapeZonationbenefit-2007}. As an example of such a function, we chose the p-norm:
\begin{equation} \label{eq: pnorm}
    \left(\sum_{i=1}^n p_i^\alpha \right)^{1/\alpha}
\end{equation}
This function weights each single species survival probability with $p_i^\alpha$, and then adds these values up. For $0< \alpha < 1$, the weighting favors an even distribution of survival probabilities, and for $\alpha = 1$ it is identical to the additive function. In a broad sense, eq. \ref{eq: pnorm} resembles the Shannon index, which is often used to express biodiversity as a function of species abundance. A summary of the three objective functions is given in Table \ref{table: functions}. 

\begin{table}[t]
  \centering
  \begin{tabular}{lll} \toprule
  \textsc{Function} & \textsc{Objective}  \\ \midrule \addlinespace[0.2cm] 
  $ \sum_{i=1}^n p_i $ & Expected number of surviving species after $T$   \\ 
  $\prod_{i=1}^n p_i$ & Probability of all species surviving after $T$     \\
  $\left(\sum_{i=1}^n p_i^\alpha \right)^{1/\alpha}$ & Sum of weighted survival probabilities\\ 
\bottomrule \\
\end{tabular}
  \caption{Overview of the analyzed objective functions.}\label{table: functions}
\end{table}

\subsection{The relation between costs and species survival}
\begin{figure}[H]
\centering
\includegraphics [width=8.4cm]{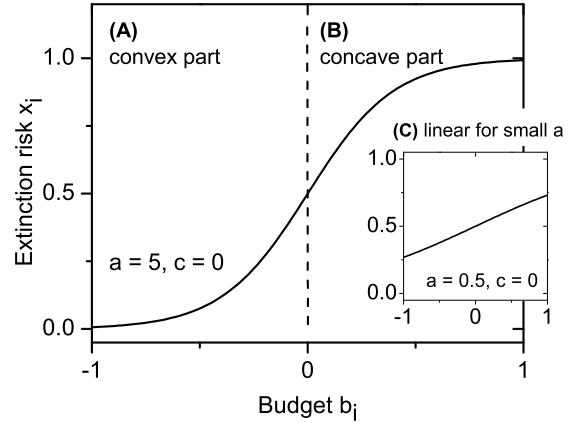}
\caption{Relationship between budget expenditure and survival of a single species for eq. \ref{eq: sigmoid} with $a=5, c=0$. A:~Below the threshold, eq.~\ref{eq: sigmoid} is convex B:~Beyond the threshold, eq.~\ref{eq: sigmoid} is concave. C:~For $a \ll B^{-1}$, eq.~\ref{eq: sigmoid} is approximately linear.} \label{figure: cost-survival figure}
\end{figure}

Ideally, the question of conservation priorities would not have to be asked, and we would simply provide each species with sufficient and adequate resources and habitat for their survival. Unfortunately, conservation is only one of many competing human ambitions. In the majority of situations, systematic conservation planning is subject to a limited budget $B$, and it has to be decided how this budget is spent most effectively \citep{Naidoo-Integratingeconomiccosts-2006, Wilson-Prioritizingglobalconservation-2006}.\\\\
This decision is further complicated because the relationship between costs spent on conservation and resulting change in population survival is often not linear. On the one hand, it is very frequently found and assumed that the costs for additional conservation increase with increasing conservation efforts \citep[see e.g.][]{Eiswerth-Maximizingconservedbiodiversity-2001, Drechsler-Combiningpopulationviability-2004, Naidoo-Integratingeconomiccosts-2006}. For example, land may get increasingly scarce and therefore more expensive when the areas used for conservation are increased \citep{Drechsler-importanceofeconomic-2001, Armsworth-FromCoverLand-2006, Polasky-Youcan'talways-2006}. On the other hand, conservation efforts often need to cross certain thresholds, such as the minimal viable population size, to become effective \citep{With-CriticalThresholdsIn-1995, Hanski-Minimumviablemetapopulation-1996, Fahrig-Howmuchhabitat-2001}. \\\\
A function which may conveniently exhibit all theses characteristics and which is therefore often used to model threshold situations is the sigmoid function (Fig. \ref{figure: cost-survival figure}). We use this function to illustrate our findings, however, all general results of this paper will not depend on the particular functional form, but only on general curvature properties of the cost-survival function. For now, let us assume an amount $b_i$ of our conservation budget $B$ will increase the annual survival rate $x_i$ of the i-th single species according to
\begin{equation}\label{eq: sigmoid}
    x_i = \frac{1}{1+e^{ -a_i \cdot (b_i + c_i)} }
\end{equation}
where $a_i$ controls the steepness of the threshold and $c_i$ represents the initial state of the species, i.e. the value which is achieved without any budget expenditures. Eq. \ref{eq: sigmoid} grows convexly (more than linearly, Fig.~\ref{figure: cost-survival figure}A) below the threshold (when $c_i + b_i< 0$) and concavely (less than linearly, Fig.~\ref{figure: cost-survival figure}B) above the threshold (when $c_i + b_i> 0$). Note that for sufficiently small steepness $a$ $(a \ll 1/B$), the cost-survival function can be considered approximately linear (Fig. \ref{figure: cost-survival figure}C), a fact that will be used in the following analysis. Furthermore, we assume that species do not interact and do not share any common resources or habitats. Thus, $x_i$ does not depend on $b_j$ with $i \neq j$. 

\subsection{The optimal conservation decision}
To compare the conservation decisions which would be made based on the discussed objective functions (eqs.~\ref{eq: additive},~\ref{eq: multiplic},~\ref{eq: pnorm}) and different time horizons $T$, we assume the following:\\\\
A landscape planner has to split a budget $B$ between two species. He spends $b_1$ on species 1 and $b_2=B-b_1$ on species 2. We call the case where most of the budget is used for one species an uneven distribution, and we call the case where the budget is spent evenly among the two species an even distribution. The annual survival probability of each species changes with $b_i$ according to eq.~\ref{eq: sigmoid}. The survival probability after the time horizon $T$ is given by eq.~\ref{eq: expdecay}. Inserting this into the three objective functions (additive, multiplicative, p-norm), we calculate the value of the objective functions (the score) for time horizons between 1 and 100 years, $b_1$ ranging from $0\%-100\%$ of the budget $B$, and different functional relationships between annual survival probability $x_i$ and budget expenditure $b_i$.

\section{Results}
Analyzing the model, it becomes evident that the effect of the time horizon depends on the relation between budget expenditures and species survival. To illustrate this, we discuss the results for four different scenarios: First, we present the results for species survival of both species depending linearly, concave (less than linearly) and convex (more than linearly) on budget expenditures. Finally, we discuss a case where the two species are in a different initial state and thus react differently to budget expenditures.  

\subsection{Linear cost-survival functions}
For species survival depending linearly on budget expenditure (as, e.g., in Fig.~\ref{figure: cost-survival figure}C), we obtain the following scores as a function of $T$ and the budget distribution: For the additive objective, we find the highest scores for uneven distributions, spending all of the budget on one of the two species. In contrast, the multiplicative objective favors an even distribution throughout all choices of the time horizon $T$. Finally, the p-norm favors an even distribution for short time horizons until a critical time $T_c$. For any $T$ larger than $T_c$, uneven distributions are favored. The results are displayed in Fig.~\ref{figure:  linear}.
\begin{figure}[H]
\begin{center}
\includegraphics [width=8.4cm]{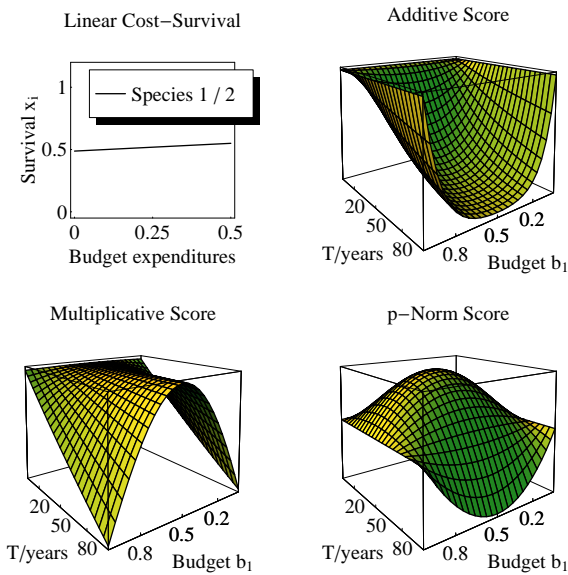}
\caption{Linear cost-survival function: The three 3-d plots show the score for the additive, the multiplicative, and the p-norm objective function. On the x-axis (right) the proportion of the budget assigned to species 1, on the y-axis (left) the time horizon $T$ in years, and on the z-axis (upwards) the score of the respective conservation option. Parameters: $a=0.4, B = 0.5, c = -0.1, \alpha = 0.013$. For each $T$, the z values are scaled to a reference value (the score that would be obtained by choosing $b_1 = 0$ for additive and p-norm functions, and $b_1 = 0.5$ for the multiplicative function) to allow the graph to be more easily read. Otherwise, cases of high $T$ would hardly be visible because survival probabilities here are naturally lower than for cases of small $T$.} \label{figure: linear}
\end{center}
\end{figure}
\subsection{Concave cost-survival functions}
\begin{figure}[H]
\centering
\includegraphics [width=8.4cm]{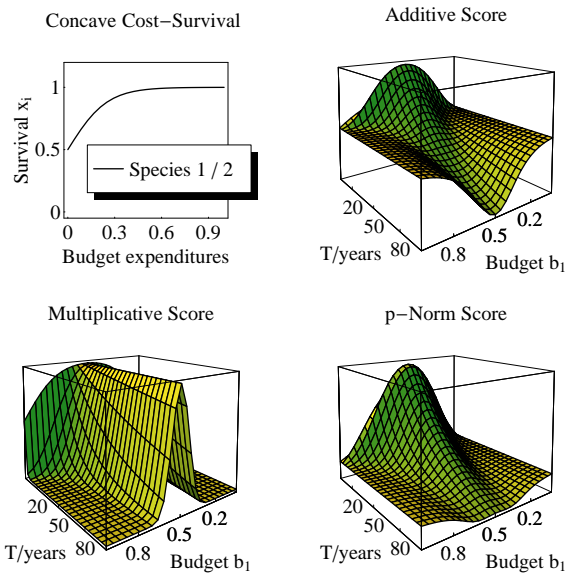}
\caption{Concave cost-survival function: The three 3-d plots show the score for the additive, the multiplicative, and the p-norm objective function. On the x-axis (right) the proportion of the budget assigned to species 1, on the y-axis (left) the time horizon $T$ in years, and on the z-axis (upwards) the score of the respective conservation option. Parameters: $a=8, B = 1, c = 0, \alpha = 0.5$. For each $T$, the z values are scaled as in Fig. \ref{figure: linear} to allow the graph to be more easily read.} \label{figure: concave}
\end{figure}

For a concave relationship between budget expenditure and annual survival probability we observe, both for the additive and the p-norm objectives, a change of conservation priorities around a critical time $T_c$. For time horizons smaller than $T_c$, an even budget distribution is favored, while at larger $T$ uneven distributions rate best. Again, the multiplicative objective favors an even distribution for all choices of $T$. The results are displayed in Fig.~\ref{figure: concave}.

\subsection{Convex cost-survival functions}
Convex cost-survival functions naturally favor uneven budget distributions, owing to the more than linear growth of survival with the budget expenditure. For moderate convexity, however, the results still resemble the linear case (Fig.~\ref{figure:  linear}) very closely. Only for a very strong convexity may the balancing influence of the multiplicative and the p-norm function eventually be overruled, and all three objectives favor an uneven distribution for any time horizon $T>1$.
\subsection{Non-even baseline values}
\begin{figure}[H]
\centering
\includegraphics [width=8.4cm]{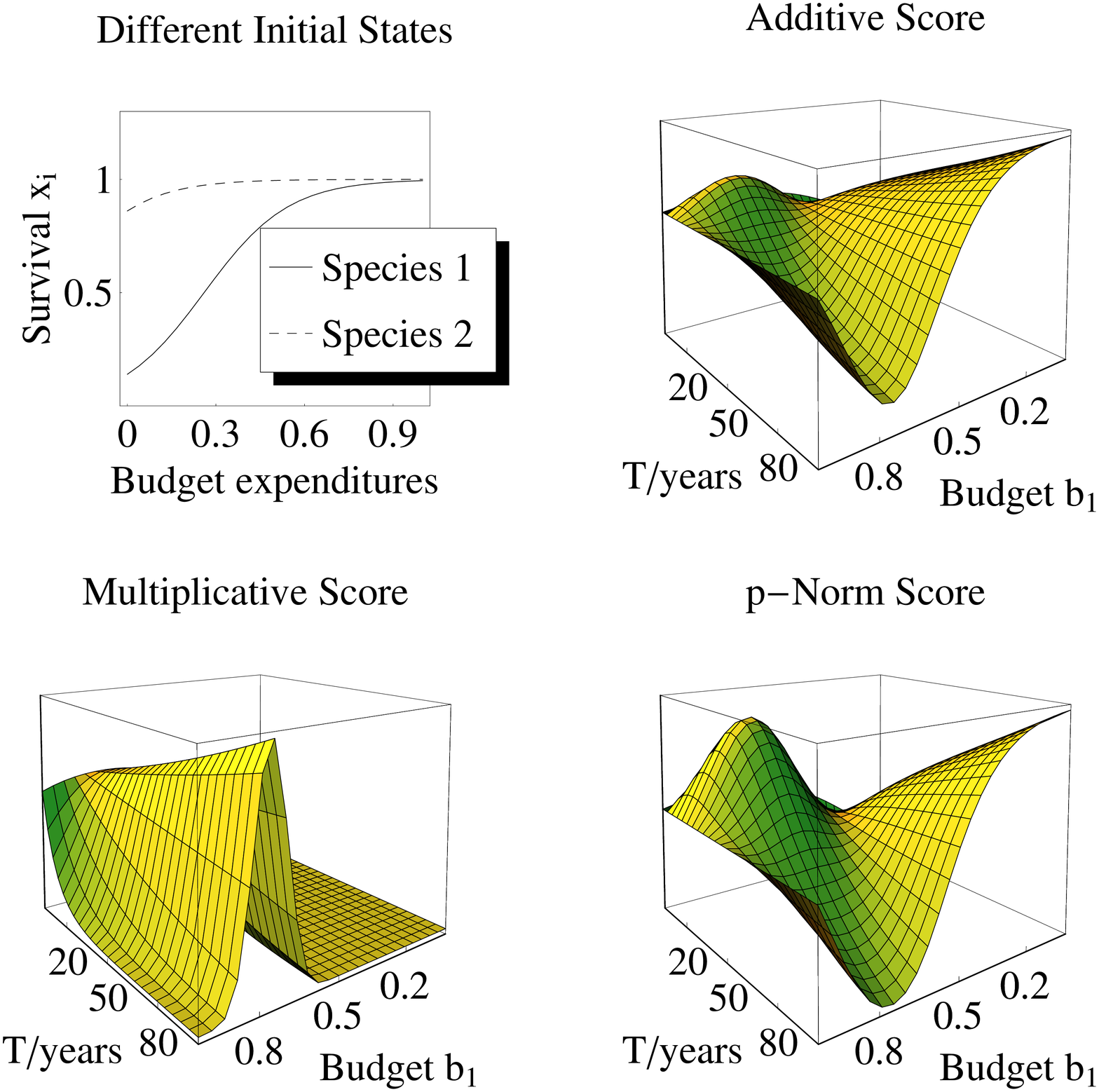}
\caption{Different initial states: The three 3-d plots show the score for the additive, the multiplicative, and the p-norm objective function. On the x-axis (right) the proportion of the budget assigned to species 1, on the y-axis (left) the time horizon $T$ in years, and on the z-axis (upwards) the score of the respective conservation option.  Parameters: $a=7, B = 1, c_1 = -0.26, c_2 = +0.26, \alpha = 0.5$. The values for the two additive objective functions are scaled as in Fig. \ref{figure: linear}, the values for the multiplicative objective are scaled at each $T$ to the value obtained by $b_1 = 0.82$.} \label{figure: uneven}
\end{figure}

Finally, we show a case with different initial states for the two species: Species 1 has a poor initial state of conservation below the threshold (convex cost-survival, see Fig.~\ref{figure: cost-survival figure}A), and species 2 is above the threshold and in a much better initial state (concave cost-survival, see Fig.~\ref{figure: cost-survival figure}B). The resulting scores are shown in Fig. \ref{figure: uneven}: Both for the additive and the p-value functions, the score favors a concentration on the threatened species 1 for short time horizons and a concentration on the more stable species 2 for long time horizons. Under a multiplicative objective, conservation budgets are always concentrated on the threatened species.

\subsection{Generalization of the results}
Are these results general, or only valid for a small or unreasonable parameter range? As we show in Appendix \ref{sec: proof time dependence}, conservation decisions with additive functions like eqs. \ref{eq: additive} and \ref{eq: pnorm} are in fact sensitive to the time horizon under quite general conditions, which is that either: a) The functional relation between budget expenditures and survival is sufficiently concave; or b) The multi-species objective function puts a sufficiently strong weight on even survival probabilities and the relationship between costs and survival is concave, linear, or sufficiently weakly convex.\\\\
Equally important, however, is whether such a sensitivity of conservation decisions will appear in real world situations. To examine the sensitivity of the model to changes in the parameters, we solved numerically for the time where conservation decisions shift between even and uneven budget distributions. The results (Appendix \ref{sec: sensitivity analysis}) show that the parameter range which yields a switch within the range of typical choices for $T$ is fairly large.\\\\
In contrast to additive objective functions, we could not find any impact of $T$ whatsoever for the case of the multiplicative function. This is no coincidence, but can easily be understood. Since the power operation commutes with the multiplication, a conservation alternative that maximizes $\prod_i x_i $ also maximizes $\prod_i p_i$ for any $T$. Therefore, a simple multiplicative function with a static budget is not influenced by the choice of the time horizon. A formal proof of this is given in Appendix \ref{sec: proof of multiplic}.

\section{Discussion} 
Evaluating multi-species survival probabilities requires the choice of an objective function which transforms survival probabilities into a single value. Different forms of objective functions have been used in the literature, some of which maximize the expected number of surviving species (additive functions), whereas others also emphasize an even distribution of survival probabilities among species (multiplicative functions or weighted additive functions). \\\\ 
Our results show that the time horizon at which species survival probabilities are calculated has a crucial impact on conservation decisions with additive functions when at least one of the following two assumptions is fulfilled: a) The functional relation between budget expenditures and survival is sufficiently concave; or b) The multi-species objective function puts a sufficiently strong weight on even survival probabilities and the relationship between costs and survival is concave, linear, or sufficiently weakly convex. For our simple case of two species, conservation decisions based on such functions change drastically when the time horizon crosses some critical value $T_c$.\\\\
The underlying reason behind this is that survival probability drops exponentially with the time horizon. While a concave cost-survival relationship or a concave objective function favor an even budget distribution for short time horizons, the exponential decay makes small differences very large in the long run and therefore eventually shifts the highest score to uneven distributions when the time horizon $T$ is increased. This time-dependence of the indicator "survival probability" constitutes a major difference to other indicators, such as expected coverage, which are used for conservation planning.\\\\
Our sensitivity analysis revealed that a crucial influence of the time horizon appears for a large range of realistic parameter combinations and functions. Therefore, a potentially drastic influence of $T$ on conservation decisions for practical cases cannot be ruled out. Only multiplicative functions showed no response to the choice of $T$ at all. This is no coincidence, but a fundamental property of multiplicative functions, as we showed. However, we do not believe that this is necessarily an argument in favor of multiplicative functions. A multiplicative function is certainly useful when the survival of all species is the main goal, but its absolute insistence on evenness can make it a dangerous choice when the budget is not large enough to conserve all species. For such cases, it may be that a distribution of the budget that maximizes a multiplicative objective minimizes the expected number of species surviving (e.g. Fig. \ref{figure: linear}).\\\\
In conclusion, we believe that the influence of time preferences on conservation decisions has not been appreciated enough in the past. This is even more so given that a lot of recent research is attracted by dynamical problems which are by their nature strongly affected by the choice of the time horizon \citep{Meir-Doesconservationplanning-2004, Drechsler-Probabilisticapproachesto-2005, McBride-Incorporatingeffectsof-2007, Pressey-Conservationplanningin-2007}. As we increasingly realize that the future challenges for conservation such as climate and global change are dynamic, time preferences will play an increasing role in conservation decisions. Thus, the time horizon must be acknowledged as a fundamental part of the objective function. It should be selected with care, and its influence should be analyzed and communicated when presenting conservation recommendations.\\\\ 
But what is the right time horizon? Ultimately, the choice of a time horizon is a normative decision. It cannot be decided on scientifically, but must be developed in interaction with stakeholders and society. To establish such an interaction, the influence of the time horizon has to be determined and openly communicated.
\section{Acknowledgements}
The authors would like to thank Silvia Wissel, Karin Johst, Volker Grimm and Atte Moilanen as well as two anonymous reviewers for helpful comments on the manuscript.

\begin{appendices}

\renewcommand{\theequation}{A.\arabic{equation}} 
\setcounter{equation}{0} 

\section{Proof of the time-dependence of additive functions}\label{sec: proof time dependence}
Assume we have two species with equal cost-survival functions. We get an even distribution as a unique solution if the summands of the objective function which are given by
\begin{equation}\label{eq. objective summands}
    \left ( x(b)\right)^{\alpha \cdot T}
\end{equation}
are concave functions of $b$ on the whole domain accessible with the budget $B$. Accordingly, we get an uneven distribution as a unique solution if eq. \ref{eq. objective summands} is convex on the whole domain. Assuming that the cost-survival function is a smooth function of $b$, all derivatives are bounded and there will be a $T_{min}$ such that eq. \ref{eq. objective summands} is concave for all $T<T_{min}$ and a $T_{max}$ such that eq. \ref{eq. objective summands} is convex for all $T>T_{max}$. Thus, the optimal budget distribution must switch or exhibit multiple solutions between $T_{min}$ and $T_{max}$. The same argument also applies for species with different cost-survival functions with the addition that optimal points may slightly shift position as can be seen in Fig. \ref{figure: uneven}.\\\\
Hence, there will always be a range of $T$ at which eq. \ref{eq. objective summands} changes from a concave to a convex function and we may observe a dramatic shift of optimal conservation decisions. For practical considerations, however, this will only be of relevance if the critical time $T_c$ where the highest score switches from even to uneven distributions, is within the range of typical choices for the time horizon $T$ (30 to 100 years). From eq. \ref{eq. objective summands}, we see directly that this can only be the case if there exists a $T$ within the considered range such that a) $x(b)$ is sufficiently concave to compensate the convex influence of $\alpha \cdot T$ in the exponent of eq. \ref{eq. objective summands} or b) $x(b)$ is concave, linear, or sufficiently weakly convex and $\alpha < 1$ is sufficiently small to make eq. \ref{eq. objective summands} linear within the considered range.

\renewcommand{\theequation}{B.\arabic{equation}} 
\setcounter{equation}{0}  

\section{Sensitivity Analysis}\label{sec: sensitivity analysis}
\begin{figure}[H]
\centering
\includegraphics [width=8.4cm]{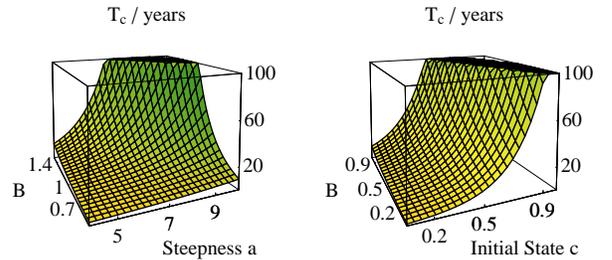}
\caption{$T_c$, the time horizon where the highest score changes from an even to an uneven budget distribution as functions of $a$, $B$ and $c$. Other parameters: $c=0$ (left panel), $a=5$ (right panel). The cost-survival function corresponding to the right panel ($a=5$) is identical with Fig.~\ref{figure: cost-survival figure}. }   \label{figure: sensitiv}
\end{figure}

To get an estimate of the sensitivity of the time $T_c$ where the budget distribution changes towards a change of parameters, let us assume we have an additive objective function, equal initial states $c_i$ and equal concave cost-survival functions. Then $T_c$ will be approximately at the time $T$ where the score of a totally uneven distribution of the budget equals the score of an even distribution:
\begin{equation} \label{eq: tc gleichsetzen}
    x(B)^{T_c}+  x(0)^{T_c}= 2 \cdot x(B/2)^{T_c} 
\end{equation} 
Here, ($B,B/2,0$) refers to the proportion of the budget $B$ to be inserted in the cost-survival function eq.~\ref{eq: sigmoid}. We solved eq.~\ref{eq: tc gleichsetzen} numerically with the sigmoid function eq. \ref{eq: sigmoid}. Fig. \ref{figure: sensitiv} shows that the range of parameters which yield times $T_c$ between $1-100$ years is fairly large.

\renewcommand{\theequation}{C.\arabic{equation}} 
\setcounter{equation}{0} 

\section{Proof of the time-independence of a multiplicative score} \label{sec: proof of multiplic}
The multiplicative score eq. \ref{eq: multiplic} can be rewritten as
\begin{equation}
   \prod_i p_i =\prod_i (x_i)^T=\left(\prod_i x_i \right)^T
\end{equation} 
As the power operations commute with the multiplication, we can factor out the power operation. The latter is strictly monotonous, hence an option which maximizes $\prod_i x_i$ also maximizes $\prod_i p_i$ for any $T$.

\end{appendices}

\bibliographystyle{D:/home/Bibliography/Databases/bibstyles/elsart-harv-hyper-arxiv}
\bibliography{D:/home/Bibliography/Databases/Flo}

\end{document}